# Splitting of chiral and deconfinement phase transitions induced by rotation

Fei Sun,[1,2,3,*] Kun Xu,[2,†] and Mei Huang[2,‡]

[1]*Department of Physics, China Three Gorges University, Yichang 443002, China*
[2]*School of Nuclear Science and Technology, University of Chinese Academy of Sciences, Beijing 100049, China*
[3]*Center for Astronomy and Space Sciences, China Three Gorges University, Yichang 443002, China*

The chiral and deconfinement phase transitions under rotation have been simultaneously investigated in the Polyakov-Nambu-Jona-Lasinio (PNJL) model. An interesting observation has been found that the chiral phase transition is catalyzed and the deconfinement phase transition is decelerated by rotation, therefore a chiral symmetric but confined phase is induced by rotation, which indicates that chiral dynamics and gluon dynamics can be split by rotation.

## I. INTRODUCTION

The quark dynamics and gluon dynamics of quantum chromodynamics (QCD) describing the internal structure of hadrons and are responsible for 99% mass of the visible matter in the universe, and the interplay between chiral and deconfinement phase transitions has been the main theme of QCD phase diagram at finite temperature and density and other extreme conditions including external strong magnetic field and vortical field.

The rotation effect is a fascinating subject of QCD matter in heavy-ion collision phenomenology nowadays. QCD matter created through noncentral heavy ion collisions carries a finite angular momentum at the order of $10^4 \sim 10^5 \hbar$ with local angular velocity in the range $0.01 \sim 0.1$ GeV [1–6], which affects both the spin and orbital angular momentum of quarks and gluons [7–12]. Experimentally, the global spin polarizations of $\Lambda$ and $\bar{\Lambda}$ have been measured by the STAR collaboration in Au + Au collisions over a wide range of beam energies $\sqrt{s_{NN}} = 7.7$–200 GeV and by ALICE collaboration in Pb + Pb collisions at 2.76 TeV and 5.02 TeV [10,13,14]. The observation of hadron polarization and spin alignment opens a new window to study the properties of quark-gluon plasma (QGP) under rotation created in Relativistic Heavy Ion Collider (RHIC) and Large Hadron Collider (LHC).

QCD matter under rotation has been studied by using effective QCD models as well as lattice simulations. It has been also attracted much interest of investigating rotation effect on other physical situations, e.g., the mesonic condensation of isospin matter with rotation in hadron physics [15], the trapped nonrelativistic bosonic cold atoms in condensed matter physics [16–20], and the rapidly spinning neutron stars in astrophysics [21–23].

Dynamical chiral symmetry breaking (DCSB) and confinement are two of the most basic features of QCD. Theoretically, the non-Abelian nature of QCD makes it difficult to have a thorough understanding of DCSB and many effective models have been proposed. In the past, much attention has been paid on the phase diagram in the plane of temperature and density. For example, The Nambu and Jona-Lasinio (NJL) model, quark-meson (QM) model, holographic QCD model, functional renormalization group (FRG), Dyson-Schwinger equations (DSE) as well as their extending models [24–47] have been used to investigate the QCD phase transition. In fact, the phase transition has been already examined extensively in the presence of finite temperature and chemical potential, in addition to the usual temperature and density affect the QCD phase transition. In recent years, much efforts have been paid on QCD phase transitions and QCD matter properties induced by rotation.

A lot of theoretical efforts have been dedicated to the study of rotational phenomena. Apart from transport properties, such as the chiral vortical effect and chiral vortical wave [48–51], it is also of significant interest to explore the effects of rotation on the phase transitions, since the rotation plays a crucial role in shaping the behavior of the phase diagram both for the chiral and deconfinement transition of QCD matter. The chiral phase transition has been explored under rotation and it is found that the chiral condensate is suppressed by the rotation [52–62].

[*]sunfei@ctgu.edu.cn
[†]xukun21@ucas.ac.cn
[‡]huangmei@ucas.ac.cn

It is natural to ask how the rotation affects another important feature of QCD, i.e., the confinement-deconfinement phase transition. There is still controversy in the literature on how rotation affects the deconfinement transition. Based on the holographic QCD approach, it is found in Ref. [63] that the deconfinement critical temperature $T_c^d$ decreases with growing angular velocity, which is confirmed by other holography studies [64–77]. Besides, the lattice QCD [78] simulation indicates that the critical temperature of the confinement/deconfinement transition in gluodynamics increases with increasing angular velocity, which is in contradiction with the results obtained in holography models.

In a hadron resonance gas model [79], the critical temperature of the confinement/deconfinement phase transition decreases with rotation, which is in agreement with the results in holographic QCD models. Therefore it calls for further studies on how rotation affects the gluodynamics and the confinement/deconfinement phase transition. In this work we will investigate the effect of rotation on chiral phase transition and confinement/deconfinement phase transition in the framework of the Polyakov-Nambu-Jona-Lasinio (PNJL) model. The PNJL model [80–86] shows features of both chiral symmetry restoration and deconfinement phase transition. So it is natural to ask what are the influences of rotation on the QCD phase structure, and how does the interplay between chiral and deconfinement phase transitions in this effective model. In this paper, we extend the PNJL model in the presence of rotation effect, and both chiral condensate and Polyakov loop modified by rotation are under considered, in particular, we are going to address the question how the rotation that in the coupling term influences the chiral transition and deconfinement transition.

Our work is organized as follows. We first give a brief description of the PNJL model and then consider the PNJL model in the presence of rotation in Sec. II, by using the mean-field approximation and the finite temperature field methods we obtain the grand thermodynamic potential. In Sec. III we present numerical results and discussions on the chiral and deconfinement phase transition under rotation. Section IV summarizes and concludes the paper.

## II. FORMALISM

Here we give a very brief sketch of the basis for studying the rotating matter. For the rotating frame, the space-time metric reads

$$g_{\mu\nu} = \begin{pmatrix} 1-\vec{v}^2 & -v_1 & -v_2 & -v_3 \\ -v_1 & -1 & 0 & 0 \\ -v_2 & 0 & -1 & 0 \\ -v_3 & 0 & 0 & -1 \end{pmatrix}, \quad (1)$$

where $v_i$ is the velocity and $v = \sqrt{v_1^2 + v_2^2 + v_3^2}$. The Lagrangian in the two-flavor NJL model under rotation can be written as follows:

$$\mathcal{L}_{\mathrm{NJL}} = \sum_f \bar{\psi}_f [i\bar{\gamma}^\mu(\partial_\mu + \Gamma_\mu) - m + \gamma^0\mu]\psi_f + G(\bar{\psi}\psi)^2, \quad (2)$$

here, $\psi$ is the quark field, $\bar{\gamma}^\mu = e_a^\mu \gamma^a$ with $e_a^\mu$ being the tetrads for spinors and $\gamma^a$ represents the gamma matrix, $\Gamma_\mu$ is defined as $\Gamma_\mu = \frac{1}{4} \times \frac{1}{2}[\gamma^a, \gamma^b]\Gamma_{ab\mu}$ which is the spinor connection, where $\Gamma_{ab\mu} = \eta_{ac}(e_\sigma^c G_{\mu\nu}^\sigma e_b^\nu - e_b^\nu \partial_\mu e_\nu^c)$, and $G_{\mu\nu}^\sigma$ is the affine connection determined by $g^{\mu\nu}$, $m$ is the bare quark mass matrix, $\mu$ denotes the chemical potential, and $G$ represents the coupling constants.

It should be noted that the NJL model lacks gluon degrees of freedom. In order to effectively account for the contribution from gluodynamics, we use the PNJL model in the following, where the Lagrangian takes the form:

$$\mathcal{L}_{\mathrm{PNJL}} = \mathcal{L}_{\mathrm{NJL}} + \bar{\psi}\gamma^\mu A_\mu \psi - \mathcal{U}(\Phi, \bar{\Phi}, T), \quad (3)$$

here, we have included the coupling between the fermion fields and the gauge fields and the effective gluonic potential $\mathcal{U}(\Phi, \bar{\Phi}, T)$, whose explicit expression will be given later.

When extending to the case of rotating fermions with nonzero chemical potential, considering a system with an angular velocity along the fixed z-axis, then $\vec{v} = \vec{\omega} \times \vec{x}$. By choosing $e_\mu^a = \delta_\mu^a + \delta_i^a \delta_\mu^0 v_i$ and $e_a^\mu = \delta_a^\mu - \delta_a^0 \delta_i^\mu v_i$ (details can be found in Refs. [6,16]), the Lagrangian can be expanded to the first order of angular velocity. Finally, the Lagrangian is given as follows:

$$\begin{aligned}\mathcal{L}_{\mathrm{PNJL}} = &\bar{\psi}[i\gamma^\mu D_\mu - m + \gamma^0 \mu \\ &+ (\gamma^0)^{-1}((\vec{\omega} \times \vec{x}) \cdot (-i\vec{\partial}) + \vec{\omega} \cdot \vec{S}_{4\times4})]\psi \\ &+ G(\bar{\psi}\psi)^2 - \mathcal{U}(\Phi[A], \bar{\Phi}[A], T),\end{aligned} \quad (4)$$

where the covariant derivative $D_\mu = \partial_\mu - iA_\mu$ determines the coupling between the Polyakov loop and quarks, $\omega$ is the angular velocity and $\vec{S}_{4\times4} = \frac{1}{2}\begin{pmatrix}\vec{\sigma} & 0 \\ 0 & \vec{\sigma}\end{pmatrix}$ is the spin operator. From the equation above, it can be observed that due to the gluon field and the rotation, the free Dirac Lagrangian has been modified. The first term corresponds to the coupling between the quark field and gluon field, the fourth term corresponds to the orbital-rotation coupling effect and the spin-rotation coupling effect. The last term in the Lagrangian represents the effective Polyakov loop potential, and the Polyakov loops $\Phi, \bar{\Phi}$ are obtained by:

$$\Phi = \frac{1}{N_c} \langle \mathrm{tr} L \rangle, \qquad \bar{\Phi} = \frac{1}{N_c} \langle \mathrm{tr} L^\dagger \rangle, \quad (5)$$

here, the Polyakov line is defined as,

$$L(\bar{x}) = \mathcal{P} \exp\left[i \int_0^\beta d\tau A_4(\bar{x}, \tau)\right], \qquad (6)$$

where $\beta = \frac{1}{T}$, $\mathcal{P}$ denotes path ordering, and $A_4 = iA_0$ is the temporal component of Eucledian gauge field $(\bar{A}, A_4)$. In this model, the quarks couple to a background (temporal) gauge field representing Polyakov loop dynamics. When performing the mean field approximation and employing the technique of path integral formulation for Grassmann variables theory, the Lagrangian is linearized to a 4-quark interaction, and the logarithm of the partition function is expressed as follows:

$$\log Z = -\frac{1}{T}\int d^3x \left(\frac{(M-m)^2}{4G}\right) + 2\log \det\frac{D^{-1}}{T} \qquad (7)$$

here the effective quark mass $M = m - 2G\langle\bar{q}q\rangle$ and $\langle\bar{q}q\rangle$ is the so-called chiral condensate.

The inverse fermion propagator $D^{-1}$ in Eq. (7) needs its trace to be taken in color, flavor, and Dirac spaces. In momentum space, it is given as follows:

$$D^{-1} = \begin{pmatrix} -i\omega_l + (n+\tfrac{1}{2})\omega + \mu - iA_4 - M & -\vec{\sigma}\cdot\vec{p} \\ \vec{\sigma}\cdot\vec{p} & -(-i\omega_l + (n+\tfrac{1}{2})\omega + \mu - iA_4) - M \end{pmatrix}, \qquad (8)$$

here we have omitted the flavor index, $\omega_l$ is Matsubara frequency, $n$ is the z-angular-momentum quantum number. The expression for $\log\det\frac{\hat{D}^{-1}}{T}$ is given as:

$$\begin{aligned}\log \det\frac{\hat{D}^{-1}}{T} &= \text{tr}\log\frac{\hat{D}^{-1}}{T} \\ &= \int d^3x \int \frac{d^3p}{(2\pi)^3} \left\langle \psi_p(x) \left| \log\frac{\hat{D}^{-1}}{T} \right| \psi_p(x) \right\rangle.\end{aligned} \qquad (9)$$

The Dirac fields can be defined in terms of the wave functions $u(x)$, $v(x)$

$$\psi_p(x) = \sum_{n,s,p}(u(x) + v(x)). \qquad (10)$$

To find solutions of the Dirac equation, we start by choosing a complete set of commuting operators consisting of $\hat{H}$, which can be obtained from Eq. (4), the momentum in the z-direction $\hat{p}_z$, the square of transverse momentum $\hat{p}_t^2$, the z-component of the total angular momentum $\hat{J}_z$ and the transverse helicity $\hat{h}_t$, here $\hat{h}_t = \gamma^5\gamma^3\vec{p}_t\cdot\vec{S}$. Note that in our calculations, we use cylindrical spatial coordinates. By solving the eigenvalue equations of the complete set of commuting operators $\{\hat{H}, \hat{p}_z, \hat{p}_t^2, \hat{J}_z, \hat{h}_t\}$, we obtain the positive and negative energy solutions of the Dirac field as follows:

$$u = \frac{1}{2}\sqrt{\frac{E+m}{E}}e^{ip_z z}e^{in\theta}\begin{pmatrix} J_n(p_t r) \\ se^{i\theta}J_{n+1}(p_t r) \\ \frac{p_z - isp_t}{E+m}J_n(p_t r) \\ \frac{-sp_z + ip_t}{E+m}e^{i\theta}J_{n+1}(p_t r) \end{pmatrix},$$

$$v = \frac{1}{2}\sqrt{\frac{E+m}{E}}e^{-ip_z z}e^{in\theta}\begin{pmatrix} \frac{p_z - isp_t}{E+m}J_n(p_t r) \\ \frac{-sp_z + ip_t}{E+m}e^{i\theta}J_{n+1}(p_t r) \\ J_n(p_t r) \\ -se^{i\theta}J_{n+1}(p_t r) \end{pmatrix} \qquad (11)$$

Here, we have chosen the overall normalization of these solutions, and $s = \pm 1$ is the transverse helicity quantum number. After the summation of all the Matsubara frequencies and the general approach of the finite temperature fields [87] is carried out, it can be shown that the thermodynamic grand potential $\Omega = -\frac{T}{V}\log Z$ takes the following form:

$$\begin{aligned}\Omega_{\text{PNJL}} = G\langle\bar{q}q\rangle^2 &- \frac{T}{4\pi^2}\sum_{n=-\infty}^\infty \int_0^\Lambda p_t dp_t \int_{-\sqrt{\Lambda^2 - p_t^2}}^{\sqrt{\Lambda^2 - p_t^2}} dp_z (J_{n+1}(p_t r)^2 + J_n(p_t r)^2) \\ &\times \text{Tr}_c[\log(1 + Le^{-\frac{\varepsilon_n - \mu}{T}}) + \log(1 + L^\dagger e^{-\frac{\varepsilon_n + \mu}{T}}) + \log(1 + L^\dagger e^{\frac{\varepsilon_n - \mu}{T}}) + \log(1 + Le^{\frac{\varepsilon_n + \mu}{T}})] + \mathcal{U}(\Phi, \bar{\Phi}, T),\end{aligned} \qquad (12)$$

finally, the grand potential reads,

$$\Omega_{\text{PNJL}} = G\langle\bar{q}q\rangle^2 - \frac{T}{4\pi^2}\sum_{n=-\infty}^{\infty}\int_0^{\Lambda} p_t dp_t \int_{-\sqrt{\Lambda^2-p_t^2}}^{\sqrt{\Lambda^2-p_t^2}} dp_z (J_{n+1}(p_t r)^2 + J_n(p_t r)^2)$$

$$\times \left[\log\left(1 + 3\Phi e^{-\frac{\varepsilon_n-\mu}{T}} + 3\bar{\Phi}e^{-2\frac{\varepsilon_n-\mu}{T}} + e^{-3\frac{\varepsilon_n-\mu}{T}}\right) + \log\left(1 + 3\bar{\Phi}e^{\frac{\varepsilon_n-\mu}{T}} + 3\Phi e^{2\frac{\varepsilon_n-\mu}{T}} + e^{3\frac{\varepsilon_n-\mu}{T}}\right)\right.$$

$$\left. + \log\left(1 + 3\bar{\Phi}e^{-\frac{\varepsilon_n+\mu}{T}} + 3\Phi e^{-2\frac{\varepsilon_n+\mu}{T}} + e^{-3\frac{\varepsilon_n+\mu}{T}}\right) + \log\left(1 + 3\Phi e^{\frac{\varepsilon_n+\mu}{T}} + 3\bar{\Phi}e^{2\frac{\varepsilon_n+\mu}{T}} + e^{3\frac{\varepsilon_n+\mu}{T}}\right)\right] + \mathcal{U}(\Phi,\bar{\Phi},T). \quad (13)$$

Here, for simplicity, we have introduced the quark quasiparticle energy $\varepsilon_n = \sqrt{M^2 + p_t^2 + p_z^2} - (\frac{1}{2}+n)\omega$ with the dynamic quark mass $M = m - 2G\langle\bar{q}q\rangle$. The Polyakov loop potential gives a deconfinement phase transition at $T = T_0$ in the pure gauge theory. The function $\mathcal{U}(\Phi,\bar{\Phi},T)$ is fixed by comparison with pure-gauge lattice QCD and reads as follows,

$$\frac{\mathcal{U}}{T^4} = -\frac{1}{2}b_2(T)\Phi\bar{\Phi} - \frac{b_3}{6}(\Phi^3 + \bar{\Phi}^3) + \frac{b_4}{4}(\Phi\bar{\Phi})^2, \quad (14)$$

where $b_2(T) = a_0 + a_1\frac{T_0}{T} + a_2(\frac{T_0}{T})^2 + a_3(\frac{T_0}{T})^3$. As a preliminary study, we restrict our analysis to the influence of the coupling term on the critical behavior, which means the Polyakov loop potential used here does not have an explicit rotational dependence. In the PNJL model, quarks couple with gauge fields through the covariant derivative [see Eq. (3)], and we will investigate the influence of coupling between quarks and gluons on the quark mass and Polyakov loop, the chiral phase transition and the deconfinement phase transition.

Then, we consider the gap equations that will be required to minimize the grand thermodynamical potential, the values are determined by solving the stationary condition, namely,

$$\frac{\partial\Omega}{\partial\langle\bar{q}q\rangle} = 0, \qquad \frac{\partial\Omega}{\partial\Phi} = 0, \qquad \frac{\partial\Omega}{\partial\bar{\Phi}} = 0. \quad (15)$$

This set of coupled equations is then solved as functions of temperature $T$, quark chemical potential $\mu$, and angular velocity $\omega$. The corresponding detailed expressions for these gap equations are listed below:

$$0 = 2G\langle\bar{q}q\rangle - \frac{3}{2\pi^2}\sum_{n=-\infty}^{\infty}\int_0^{\Lambda} p_t dp_t \int_{-\sqrt{\Lambda^2-p_t^2}}^{\sqrt{\Lambda^2-p_t^2}} dp_z (J_{n+1}(p_t r)^2 + J_n(p_t r)^2)$$

$$\times \left(1 - \frac{1 + 2e^{\frac{\mu+\varepsilon_n}{T}}\Phi + e^{\frac{2(\mu+\varepsilon_n)}{T}}\bar{\Phi}}{1 + e^{\frac{3(\mu+\varepsilon_n)}{T}} + 3e^{\frac{\mu+\varepsilon_n}{T}}\Phi + 3e^{\frac{2(\mu+\varepsilon_n)}{T}}\bar{\Phi}} - \frac{1 + e^{\frac{2(\varepsilon_n-\mu)}{T}}\Phi + 2e^{\frac{\varepsilon_n-\mu}{T}}\bar{\Phi}}{1 + e^{\frac{3(\varepsilon_n-\mu)}{T}} + 3e^{\frac{2(\varepsilon_n-\mu)}{T}}\Phi + 3e^{\frac{\varepsilon_n-\mu}{T}}\bar{\Phi}}\right)\left(-\frac{2GM}{\sqrt{p_t^2 + p_z^2 + M^2}}\right), \quad (16)$$

$$0 = -\frac{3}{2\pi^2}T\sum_{n=-\infty}^{\infty}\int_0^{\Lambda} p_t dp_t \int_{-\sqrt{\Lambda^2-p_t^2}}^{\sqrt{\Lambda^2-p_t^2}} dp_z \left[(J_{n+1}(p_t r)^2 + J_n(p_t r)^2)\right.$$

$$\left.\times e^{\frac{\mu+\varepsilon_n}{T}}\left(\frac{1}{e^{\frac{3\mu-\varepsilon_n}{T}} + e^{\frac{2\varepsilon_n}{T}} + 3e^{\frac{\mu+\varepsilon_n}{T}}\Phi + 3e^{\frac{2\mu}{T}}\bar{\Phi}} + \frac{1}{1 + e^{\frac{3(\mu+\varepsilon_n)}{T}} + 3e^{\frac{\mu+\varepsilon_n}{T}}\Phi + 3e^{\frac{2(\mu+\varepsilon_n)}{T}}\bar{\Phi}}\right)\right]$$

$$+ T^4\left(-\frac{1}{2}b_3\Phi^2 + \frac{1}{2}b_4\Phi\bar{\Phi}^2 - \frac{1}{2}b_2(T)\bar{\Phi}\right), \quad (17)$$

$$0 = -\frac{3}{2\pi^2}T\sum_{n=-\infty}^{\infty}\int_0^{\Lambda} p_t dp_t \int_{-\sqrt{\Lambda^2-p_t^2}}^{\sqrt{\Lambda^2-p_t^2}} dp_z \left[(J_{n+1}(p_t r)^2 + J_n(p_t r)^2)\right.$$

$$\left.\times e^{\frac{2\mu+\varepsilon_n}{T}}\left(\frac{e^{\frac{\varepsilon_n}{T}}}{1 + e^{\frac{3(\mu+\varepsilon_n)}{T}} + 3e^{\frac{\mu+\varepsilon_n}{T}}\Phi + 3e^{\frac{2(\mu+\varepsilon_n)}{T}}\bar{\Phi}} + \frac{1}{e^{\frac{3\mu}{T}} + e^{\frac{3\varepsilon_n}{T}} + 3e^{\frac{\mu+2\varepsilon_n}{T}}\Phi + 3e^{\frac{2\mu+\varepsilon_n}{T}}\bar{\Phi}}\right)\right]$$

$$+ T^4\left(-\frac{1}{2}b_3\bar{\Phi}^2 + \frac{1}{2}b_4\Phi^2\bar{\Phi} - \frac{1}{2}b_2(T)\Phi\right). \quad (18)$$

TABLE I. Parameters of the Polyakov loop sector of the model.

| $a_0$ | $a_1$ | $a_2$ | $a_3$ | $b_3$ | $b_4$ | $T_0$ |
|---|---|---|---|---|---|---|
| 6.75 | −1.95 | 2.625 | −7.44 | 0.75 | 7.5 | 0.27 GeV |

## III. NUMERICAL RESULTS AND DISCUSSIONS

In this section, we present our numerical results for the PNJL model under rotation. For the Fermionic sector, the parameters are chosen as $m = 0.005$ GeV, $\Lambda = 0.65$ GeV, and $G = 4.93$ GeV$^{-2}$, as reported in Ref. [24], which are fixed to reproduce the physical observations. For the Polyakov loop sector, we list the parameters in Table I, which are taken from Ref. [85]. Additionally, the $z$-angular-momentum quantum number is denoted by $n = 0, \pm 1, \pm 2 \ldots$. In principle, we should sum over all the values of $n$; fortunately, these expressions converge rapidly, allowing us to limit the sum over $n$ from −5 to 5. We set the radius to $r = 0.1$ GeV$^{-1}$ and ensure that $\omega r < 1$ in all calculations.

The quark condensate, related to the dynamical quark mass, is often considered as an order parameter for spontaneous chiral symmetry breaking, while the Polyakov loop serves as an order parameter for the confinement-deconfinement phase transition. In Fig. 1, we present the evolution of the light quark effective mass and the Polyakov loop at $\mu = 0$ for different angular velocities as functions of temperature in Figs. 1(a) and 1(b), respectively. It can be observed that the effective mass of the light quark decreases with increasing temperature, indicating the restoration of chiral symmetry at high temperatures. The dynamical quark mass is seen to be suppressed by the angular velocity. On the other hand, the Polyakov loop increases as the temperature rises, showing a behavior similar to the general PNJL model. Notably, the quark mass is strongly affected by the presence of angular velocity at low temperatures, while the Polyakov loop is less affected, especially compared to the quark mass. The restoration of chiral symmetry is realized at rapid rotation in the low-temperature region. A transition from the confinement regime at low temperatures to the deconfinement regime at high temperatures is evident, and the rotation effect enhances the Polyakov loop at low temperatures but suppresses it at high temperatures. It can be observed that the Polyakov loop's dependence on rotation is mild when only considering the coupling term contribution.

Next, we consider a nonzero chemical potential where $\mu = 0.1$ GeV, and the results of light quark dynamical mass and the Polyakov loop $\Phi$ and $\bar{\Phi}$ are presented in Figs. 2(a) and 2(b), respectively. Comparing with Fig. 1(a), we observe that in the presence of a finite quark chemical potential, the light quark dynamical mass curves become smoother for all different angular velocities, indicating that the quark chemical potential suppresses the chiral condensate in the rotating system. Furthermore, we find that $\Phi$ and $\bar{\Phi}$ differ from each other due to the quark chemical potential. Although the Polyakov loop and its conjugate have some small quantitative differences for the same angular velocity, they both exhibit the same qualitative behavior, and at high temperatures, they almost coincide again.

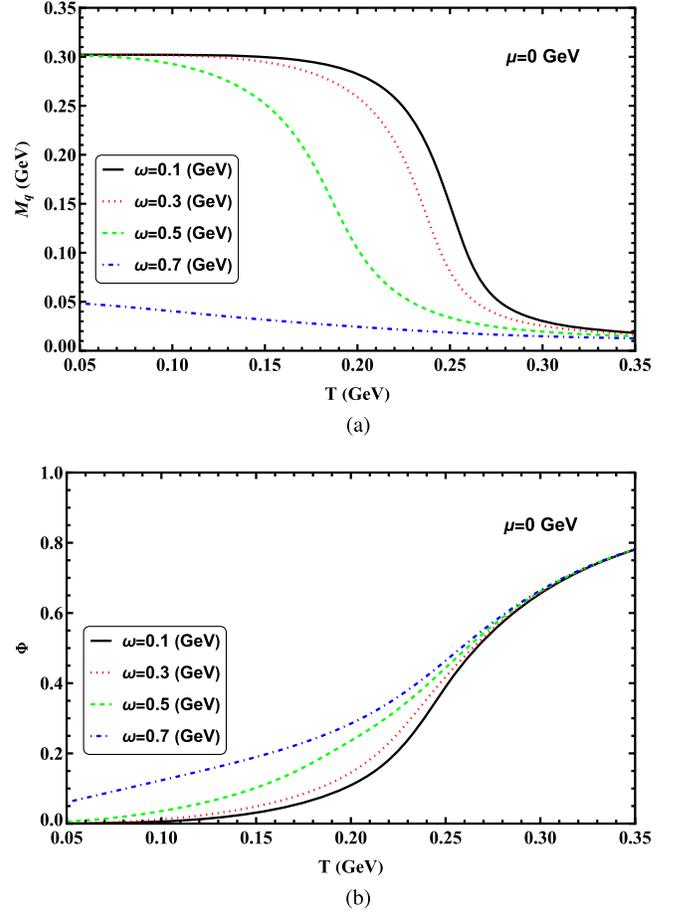

FIG. 1. The light quark effective mass and Polyakov loop as functions of temperature $T$ at $\mu = 0$ GeV for different angular velocities.

When including the quark field, the center symmetry is explicitly broken, which means there is no strict order parameter. However, the Polyakov loop can be considered as an approximate order parameter and serves as an indicator of a rapid crossover toward the deconfinement transition. Pseudocritical temperatures of the quark and Polyakov loop at zero chemical potential are depicted in Fig. 3(a). We observe that the pseudocritical temperatures of the Polyakov loop are only mildly dependent on the angular velocity compared to those of the light quark. In the small angular velocity region, there is a slowly declining trend for the pseudocritical temperatures of the Polyakov loop. As the angular velocity increases, the crossover transitions for the chiral condensate and the Polyakov loop coincide at $T_{pc} \simeq 0.235$ GeV and $\omega \simeq 0.3$ GeV. After reaching this point, the pseudocritical temperature experiences a sharp jump, and the pseudocritical temperatures of the Polyakov loop moderately increase with increasing angular velocity. In the mid-range of angular velocity, the data show a jumplike feature, the origin of which is not completely understood, but it probably indicates some form of competition between temperature and rotation. To better understand how rotation influences a phase transition, we plot $d\Phi/dT$ as a function of temperature at zero chemical potential for different angular velocities in Fig. 3(b). In this figure, the definition of $T_{pc}$ is determined by the maximum of the derivative of $\Phi$ with respect to $T$. The figure shows that, initially, the peak position moves to a lower temperature as the angular velocity increases, and then moves to a higher temperature with further increase in angular velocity. Moreover, we find that the angular velocity increases as the peak height becomes lower. It appears that the maximum of the black dashed curve is flattening, which means the Polyakov loop linearly changes with the angular velocity in this region. Once exceeding this region, there is a jump in the pseudocritical temperatures, reflecting the competition between temperature and angular velocity. Thus, the present approach suggests that the coupling term has a soft influence on the Polyakov loop, and the behavior of the critical temperature for confinement is nonmonotonic.

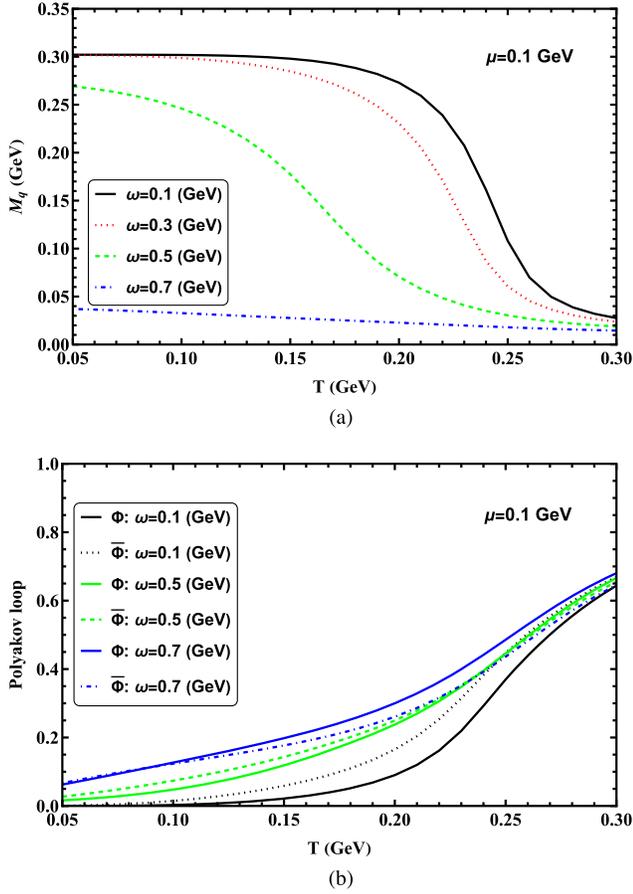

FIG. 2. The light quark effective mass and the Polyakov loop as functions of temperature $T$ at $\mu = 0.1$ GeV for different angular velocities.

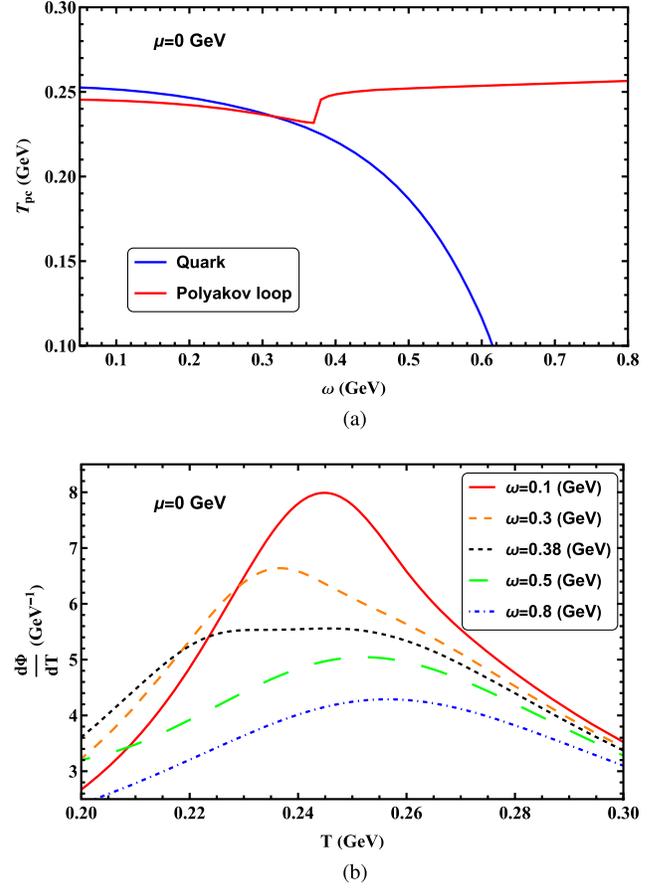

FIG. 3. The pseudocritical temperatures of the dynamical quark mass and Polyakov loop according to the angular velocity at zero chemical potential (a). Susceptibilities $d\Phi/dT$ as a function of temperature at zero chemical potential for different angular velocities (b).

From Fig. 3, it is evident that the deconfinement phase transition occurs at lower temperatures compared to the chiral phase transition in the region of small angular velocity. However, this is not a universal property of the transition under rotation. It should be emphasized that the value of $T_0$ can be rescaled over a large range. When we change $T_0$ to 0.32 GeV, a perfect coincidence of the chiral and deconfinement transitions is achieved at $\mu = 0$, $\omega = 0$, but the critical temperatures are shifted relative to each other by several tens of MeV at finite $\omega$ as shown in Fig. 4 clearly, indicating that a chiral symmetric but confined phase can be induced by rotation, and the splitting between the two transitions becomes wider with increasing angular velocity. Thus, the rotation effect, arising from the coupling between quarks and the Polyakov loop (or the contribution from rotating quarks to the Polyakov loop), has a strong impact on the splitting of the chiral transition and deconfinement transition.

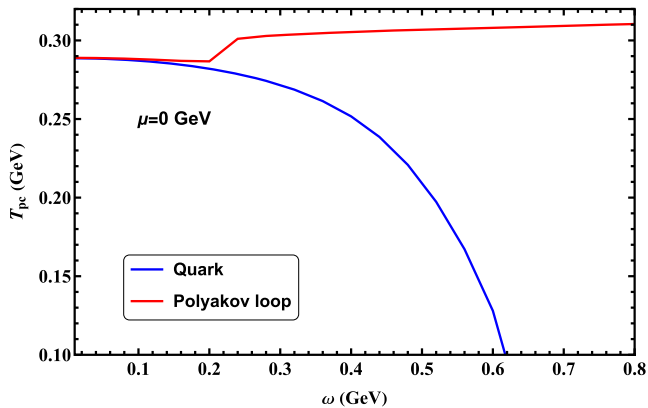

FIG. 4. The pseudocritical temperatures of the dynamical quark mass and Polyakov loop according to the angular velocity at zero chemical potential with $T_0 = 0.32$ GeV.

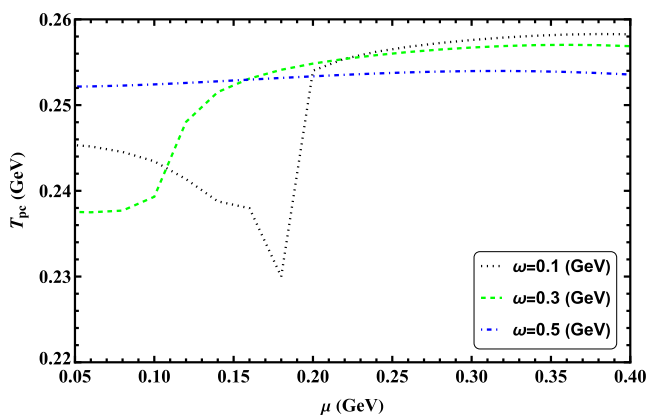

FIG. 5. The pseudocritical temperatures of the Polyakov loop $\Phi$ according to the quark chemical potential for different angular velocities with $T_0 = 0.27$ GeV.

It would be intriguing to investigate the dependence of the deconfinement transition on both quark chemical potential and rotation. Figure 5 displays the pseudocritical temperatures of the Polyakov loop $\Phi$ at different angular velocities as a function of quark chemical potential. Here, we fix $T_0 = 0.27$ GeV and vary the quark chemical potential for three different values of $\omega = 0.1$, 0.3, 0.5 GeV. When the angular velocity is small, the pseudocritical temperature initially decreases and then increases with increasing quark chemical potential. However, for sufficiently large angular velocities, the kink in the curve disappears, and at very high angular velocities, the pseudocritical temperature is almost unaffected by the quark chemical potential.

## IV. CONCLUSIONS

This work aims to develop a general model suitable for studying confinement and chiral symmetry breaking under rotation. By combining the NJL model under rotation with the Polyakov loop, we have formulated and explored QCD matter under rotation using the Polyakov loop-extended NJL model. We derived explicit analytical expressions for the PNJL model under rotation and provided detailed calculation procedures. From a phenomenological perspective, this framework offers a possibility for the theoretical study of QCD matter under rotation.

In this article, we focus on the rotational effect of the coupling between quarks and the gauge field on the chiral transition and deconfinement transition. Our findings indicate that rotation plays a significant role in the chiral transition. At low temperatures and small angular velocity, chiral symmetry is spontaneously broken, and as the angular velocity increases, the chiral symmetry gradually restores. As for the Polyakov loop, at low-temperature regions, its magnitude is enhanced by rotation, whereas at high-temperature regions, the magnitude is suppressed. Our calculations provide a physical picture of the transition under rotation, showing that the deconfinement phase transition is insensitive to the rotating effect compared to the chiral transition. We also find that rotation decreases the critical temperature of the chiral transition and increases the critical temperature of the deconfinement transition at large angular regions. Additionally, our calculations suggest that the rotation effect may induce the splitting of chiral and deconfinement phase transition thus leads to a chiral symmetric but confined phase and has a strong impact on the splitting of the chiral phase transition and deconfinement phase transition, which arises from the coupling of the chiral order parameter to the Polyakov loop.

It should be noted that, for simplicity, we did not consider the boundary effect of the system. Since any uniformly rotating system should be spatially bounded, the boundaries could modify the properties of the rotating system [52,55,56,64,77]. In addition, In Ref. [77], an inhomogeneity of plasmas was predicted due to the Tolman-Ehrenfest effect, so the inhomogeneity should also be regarded. We leave the finite volume boundary effect and the inhomogeneity effect for further study. Furthermore, the vector interactions [88–90] may play an important role in the chiral transition of the PNJL model in the presence of rotation. Additionally, the rotational influence on the equation of state may also be intriguing and requires further investigation.

We would like to point out that in the PNJL model, the insensitivity of the deconfinement phase transition under rotation, which is similar to that of at finite chemical potential. This behavior might be caused by the model itself, where the thermodynamic effect of gluodynamics is introduced through the Polyakov loop potential. In order to fully understand the gluodynamics under rotation, we may need to consider the dynamical gluons as a vector field, which should be sensitive to the rotation like the vector meson field under rotation as shown in [91]. Therefore, the behavior of full QCD matter under rotation remains an open question, and the effect of rotation on the deconfinement phase transition is still highly debated and warrants further research. In essence, the behavior of rotating QCD matter may be quite intricate. A clearer picture may emerge as more experimental data on QCD matter under rotation are accumulated. Furthermore, some recent lattice QCD calculations [92–96] show that the solidly rotating gluon plasma experiences instability, which demonstrate that the behavior of rotating QCD is complicated, and Ref. [93] demonstrates analytically that this effect is linked to the thermal melting of the magnetic part of the gluonic condensate, which gives a negative moment of inertia to the gluonic component of the plasma. So, considering the contribution of the dynamical gluon from further investigations in the effective model would be helpful. As research progresses, we believe that our understanding of this issue will deepen, providing more clues to the unconventional properties of the QCD matter under rotation.

## ACKNOWLEDGMENTS

We would like to thank Maxim Chernodub, Jinfeng Liao, Xuguang Huang, Anping Huang, Jie Mei, Rui Wen, Shijun Mao for useful discussions. The work has been supported by the National Natural Science Foundation of China (NSFC) with Grant No. 12235016, No. 12221005 and the Strategic Priority Research Program of Chinese Academy of Sciences under Grant No. XDB34030000, the start-up funding from University of Chinese Academy of Sciences (UCAS), and the Fundamental Research Funds for the Central Universities, and the Science Research Foundation of China Three Gorges University with Grant No. KJ2015A007.